\documentclass[traditabstract]{aa} 
\usepackage{graphicx}
\usepackage{txfonts}
\begin{document}
   \title{Close binary and other variable stars in the solar-age Galactic open cluster M67 \thanks{Based on observations made with the Mercator Telescope, operated on the island of La Palma by the Flemish Community, at the Spanish Observatorio del Roque de los Muchachos of the Instituto de Astrofísica de Canarias and T\"UB\.ITAK National Observatory Russian-Turkish telescope operated at  Sakl{\i}kent, Antalya.}}
  \titlerunning{Binary and other variable stars in the open cluster M67}

\author{K.~Yakut\inst{1,2,3,}\thanks{E-Mail: yakut@ast.cam.ac.uk} \and W.~Zima\inst{1} \and B.~Kalomeni\inst{2,4} \and H.~Van~Winckel\inst{1} \and C.~Waelkens\inst{1} \and P. De Cat\inst{1,5} \and
E.~Bauwens\inst{1} \and M.~Vu${\check{\rm{c}}}$kovi${\acute{\rm{c}}}$\inst{1} \and S.~Saesen\inst{1} \and L.~Le~Guillou\inst{1,6} \and M.~Parmaks{\i}zo\u{g}lu\inst{7} \and K.~Ulu\c{c}\inst{7} \and
I.~Khamitov\inst{7} \and G.~Raskin\inst{1} \and  C.~Aerts\inst{1,8}
}
   \institute{
Instituut voor Sterrenkunde, Katholieke Universiteit Leuven, Celestijnenlaan 200D, B-3001 Leuven, Belgium
 \and
University of Cambridge, Institute of Astronomy, Madingley Road, Cambridge CB3 0HA, UK
 \and
University of Ege, Department of Astronomy \& Space Sciences, 35100  \.Izmir,  Turkey
 \and
\.Izmir Institute of Technology, Department of Physics, 35430 \.Izmir, Turkey
 \and
Royal Observatory of Belgium, Ringlaan 3, 1180 Brussel, Belgium
 \and
LPNHE - Universit\'{e} Paris, 75252 Paris Cedex 05, France
 \and
T\"UB\.ITAK National Observatory, Antalya, Turkey
 \and
IMAPP, Radboud Universiteit Nijmegen, P.O. Box 9010, 6500 GL Nijmegen, the Netherlands
  }

   \date{Received aaaa xx, xxxx; accepted aaaa xx, xxxx}


\abstract{ We present  multi-colour time-series CCD photometry of the solar-age
galactic open cluster M67 (NGC~2682). About 3600 frames spread over 28
nights were obtained with  the 1.5 m Russian-Turkish and 1.2 m
Mercator telescopes. High-precision observations of the close
binary stars  AH Cnc, EV Cnc, ES Cnc, the $\delta$ Scuti type systems EX
Cnc and EW Cnc, and some long-period variables belonging to M67 are
presented. Three full multi-colour light curves of the overcontact
binary AH Cnc were obtained during three observing seasons. Likewise we gathered
three light curves of EV Cnc, an EB-type binary, and two
light curves of ES Cnc, a blue straggler binary. Parts
of the light change of  long-term variables S1024, S1040, S1045,
S1063, S1242, and S1264 are obtained.
Period variation analysis of AH Cnc, EV Cnc, and ES Cnc were done using all times of mid-eclipse available in the literature and those obtained in this study. In addition, we analyzed multi-colour light curves of the close binaries and also determined new frequencies for the $\delta$ Scuti systems. The physical parameters of the close binary
stars were determined with simultaneous solutions of multi-colour light and radial velocity curves. Finally we determined the distance of M67 as 857(33) pc via binary star parameters, which is consistent with an independent method from earlier studies.
               }

\authorrunning{K.Yakut et al.} 
   \maketitle
%

\section{Introduction}
The open cluster M67 (NGC 2682) is located at a distance of $\sim$840
pc and its stellar content consists of stars of solar age and
composition. Van den Berg \& Stetson (2004) give an isochrone age of 4.0 Gyr using
the working hypothesis that isochrones that treat overshooting are
able to reproduce the morphology of the cluster turnoff.
Hurley et al. (2005) present a direct N-body model of M67 from zero to 4 Gyrs
age. The authors not only take the cluster dynamics into account
but also the evolution of stars and binary systems. The N-body
model's initial setup for a given cluster is a statistical
realisation where the mass, the position, and the velocity of a star
are deduced from a distribution function. The model of Hurley et al. (2005)
contains 20 blue stragglers at 4 Gyr, nine of which are in binaries.
Tian et al. (2006) present simulations for the primordial blue
stragglers belonging to M67. Four of them had been
predicted by their simulations using appropriate parameters.

Besides evolved red giants, the HR diagram of M67 contains main-sequence stars.
The mass of the stars that reside
near the turnoff point provides a crucial astrophysical tool for
testing the predictions of stellar evolution models. Several clusters
show that a group of hot stars known as blue stragglers still appear to be on the main sequence above the
turnoff point. These stars are therefore hotter and bluer than other
cluster member stars with the same luminosity. They may be regarded
as unique astrophysical objects in the study of stellar evolution
theories since they show an evolution stage unexpected by standard
theories. The current scenarios for explaining the blue stragglers are
(1) post-mass transfer binaries, (2) merged former binaries, or (3)
the products of dynamical collisions. All of these models produce
rejuvenated main sequence stars.
The merging of two stars would create a hot, massive, and more
luminous star with the same age; hence, blue straggler stars would have more hydrogen in their
cores, making them appear much younger star than they
are. This hypothesis can in principle be tested by studying the pulsation properties of blue
straggler stars whose asteroseismological properties are expected to be
different from those of normal pulsating stars with similar masses and
luminosities.

The solar age cluster M67 is expected to contain active stars
similar to our Sun, and some binaries belonging to the cluster (e.g.
ES Cnc) are found to be chromospherically active stars based on the
observed Ca~II H-K lines in their spectrum. For instance, AG Cnc
(S113) (Kaluzny \& Radczynska, 1991) was found to be a
low-temperature contact (WUMa-type) binary. Recently, Giampapa et
al. (2006) studied the chromospheric activity of solar-like stars
in M67. The authors studied Ca~II  H and K lines of 60 stars.
Giampapa et al. (2006) predict the age of the open cluster M67 to
be on the range 3.8 -- 4.3 Gyr. In the cluster M67, FK Com type
giant stars and white dwarfs are also found. Landsman et al. (1998)
reported that the number of white dwarfs (WD) in M67 is consistent
with those expected from WD cooling models. X-ray observations of
the cluster reveal somewhat special stars of the cluster. ROSAT
X-ray observations of M67 indicate that active synchronized binaries
have rapid rotation despite their rather advanced ages. A magnetic
cataclysmic variable star EU Cnc, due its lower visual magnitude, is
identified with ROSAT observations. The photometric variable stars S1063
and S113 belonging to M67, independently identified as X-ray
sources by ROSAT (Belloni, Verbunt \& Schmitt 1993; Belloni et al.
1998), are also reported to be a spectroscopic binary by Mathieu et
al. (2003).

AH~Cnc ($V_{\textrm{mag}}=13.33$, \textsc{F7V}) is a
low-temperature contact binary (LTCB) system. The variability of AH
Cnc has been reported by Kurochkin (1960) who classified it as an RR
Lyrae type variable. Later on, the system was classified as WUMa-type by
Efremov et al. (1964) and Kurochkin (1965). Though the system has
been studied photometrically by numerous researchers, the radial
velocity curve of the system was obtained only by Whelan et al.
(1979). The mass ratio obtained from a spectroscopic study of the system displays
a difference from those obtained by photometric studies. Previous modelling of
the system led to orbital and physical parameters and was based on single colour RTT150 data (Yakut \& Aerts 2006).

EV Cnc was discovered to be a contact binary by Gilliland et al.
(1991). The reported period of the system is 0.44125 days (ibid).
van den Berg et al. (2002) published the light curve of the contact
binary through partial eclipses and ellipsoidal variations due to
tidal effects of the components. The observed X-ray emission from
the system is believed to be due to the coronal activity of the
magnetically active component. The observed small amplitude light
variation is due to the small inclination of the system or to the
high mass ratio of the component stars (van den Berg et al. 2002).
WUMa-type light curve properties related with unequal eclipses, were visible in
the light curve of van den Berg et al. (2002), i.e., at phase
0.25 the system is brighter than it is at phase 0.75. This feature has been
explained with a hot spot activity on the secondary star. Blake \&
Rucinski (2004) tested a technique of combining Fourier coefficients
of the light curves with spectroscopically obtained mass ratios to
provide distance calculations to the binary. However, EV Cnc was
found to be unsuitable for such distance determination due to its
faintness.

ES Cnc (S1082) is classified as a member of the RS CVn type binaries,
which are detached and active systems. Their evolution is mainly driven by
angular momentum loss by a stellar wind. ES Cnc (S1082), with its orbital
period of 1.08 days, falls in the blue part of the colour-magnitude (CM) diagram. ES Cnc is an unusual
partial-eclipsing blue straggler star. Its light
curve obtained by Sandquist \& Shetrone (2001) was
reported to show variations on timescales of a month and shorter.
Sandquist et al. (2003) showed the variation of the system in the V band to be in the range 0.01-0.03 mag
from month to month. Such variations are also detected in our new
observations of the binary. Sandquist et al. (2003) explained
them as due to spots on the cold and faint component star.
The authors also showed the existence of a
third body orbiting the binary in a highly eccentric ($e$ = 0.6)
orbit with a 1189 days period. The primary and the third components
are blue stragglers (Sandquist \& Shetrone 2001). The stars were
identified from their systematic velocities as members of M67. The
spectroscopic detection of a binary companion to the blue straggler S1082
was confirmed from high-resolution spectra. The measured
projected rotational speed of the component is 90$\pm$20 km~s$^{-1}$
and its radial velocity was found to vary with a peak-to-peak amplitude
of $\leq 25$ km s$^{-1}$. The radial velocity period of the
system does not correspond with the period derived from partial eclipses.
Therefore, it is believed that the system undergoes mass transfer (Shetrone \&
Sandquist 2000). The V-band light curve of the system shows brightness
variations due to spot activity of the secondary cool component which is nearly
synchronized (Sandquist et al.\ 2003).

Gilliland et al. (1991) studied for the first time the $\delta$ Scuti
type pulsating properties of EW Cnc  and EX Cnc,
members of M67. Gilliland \& Brown (1992) detected 10 independent modes for EW Cnc and 5
modes for EX Cnc. Recently, these stars were studied by Sandquist \&
Shetrone (2003) and Zhang et al. (2005). Zhang et al. (2005)
detected 4 modes for EW Cnc and 5 modes for EX Cnc. A spectroscopic
study was performed by Milone \& Latham (1992) and they showed that
the system is indeed a binary with a period of 4.2 days and an
eccentricity $e=0.2$.

\section{New observations, data reduction, and new light curves}
CCD observations of M67 were obtained with the
1.5 m Russian-Turkish telescope (RTT150) at T\"UB\.ITAK National
Observatory (TUG, Antalya, Turkey) and with the 1.2 m Mercator
telescope at Roque de los Muchachos Observatory on La Palma Island
(Canary Islands, Spain). The cluster has been observed on 5 nights in 2005 with the 1.5 m telescope and on 23 nights in 2006 with the 1.2 m telescope. The journal of the observations is given in Table~\ref{m67tab1}.

TUG observations were obtained with the ANDOR CCD on the RTT150. The CCD
chip has 2K$\times$2K pixels and the corresponding field of view is 8.2$^{\prime}$ x 8.2$^{\prime}$. The gain corresponds to 1.2 $e^-$/ADU
with readout noise of 2 electrons rms. During the observations a
Cousins R$_{\textrm{c}}$ (670 nm) filter was used. The 2K$\times$2K MEROPE CCD
camera was used during observations with the Mercator telescope. The
gain corresponds to 0.93 $e^-$/ADU with a readout noise of 4.5
electrons rms and the corresponding field of view is 6.5$^{\prime}$ x 6.5$^{\prime}$. The V (VG-548.26 nm), R (RG-662.91 nm), and I (IC-749.18 nm)  filters were used.
The exposure times were 15 s for V, 7.5 s for R, 10 s for I and, 2 s for R$_\textrm{c}$ filter.

\begin{figure}
\centering
\includegraphics{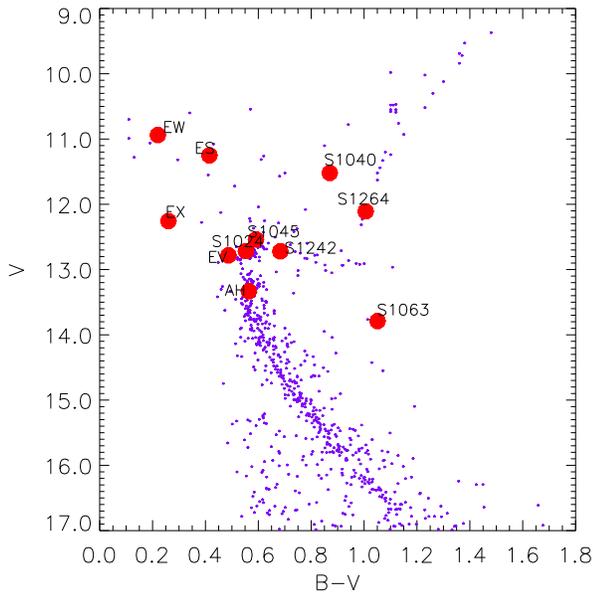}
  \caption{The colour magnitude diagram of open
cluster M67. The colours and the magnitudes of the stars are obtained
from Montgomery, Marschall \& Janes (1993).}\label{11918fg1}
\end{figure}

\begin{table}
\caption{Summary of the observations of M67. JD$^*$ lists JD
start+2453000}\vspace{0.25cm} \label{m67tab1}
\begin{tabular}{llllcc}
\hline
Run     &   UT Date           & JD$^*$ & Telescope  &  Filter & N$_{\rm{Obs}}$     \\
\hline
1       &   2005 February 23  & 425.36-425.50     &RTT150      &R$_\textrm{c}$    &   270   \\
2       &   2005 February 24  & 426.26-426.58     &RTT150      &R$_\textrm{c}$    &   464   \\
3       &   2005 April    28  & 489.26-429.40     &RTT150      &R$_\textrm{c}$    &   205   \\
4       &   2005 November  7  & 482.36-482.66     &RTT150      &R$_\textrm{c}$    &   166   \\
5       &   2005 November  8  & 683.50-683.66     &RTT150      &R$_\textrm{c}$    &   284   \\
6       &   2006 January   7  & 743.57-743.71     &Mercator    &VRI  &   288   \\
7       &   2006 January  14  & 750.56-750.80     &Mercator    &VRI  &   504   \\
8       &   2006 January  19  & 755.42-755.49     &Mercator    &VRI  &   102   \\
9       &   2006 April    29  & 855.39-855.47     &Mercator    &VRI  &   18    \\
10      &   2006 May       3  & 859.40-859.46     &Mercator    &VRI  &   123   \\
11      &   2006 May       4  & 860.39-860.47     &Mercator    &VRI  &   143   \\
12      &   2006 May       5  & 861.39-861.47     &Mercator    &VRI  &   147   \\
13      &   2006 May       6  & 862.40-862.51     &Mercator    &VRI  &   123   \\
14      &   2006 May       7  & 863.39-863.47     &Mercator    &VRI &   147   \\
15      &   2006 May       8  & 864.41-864.47     &Mercator    &VRI  &   69    \\
16      &   2006 May       9  & 865.40-865.47     &Mercator    &VRI  &   69    \\
17      &   2006 May      10  & 866.39-866.46     &Mercator    &VRI  &   72    \\
18      &   2006 May      11  & 867.40-867.46     &Mercator    &VRI  &   63    \\
19      &   2006 May      12  & 868.39-868.45     &Mercator    &VRI  &   66    \\
20      &   2006 May      13  & 869.41-869.45     &Mercator    &VRI  &   54    \\
21      &   2006 May      14  & 870.41-870.46     &Mercator    &VRI  &   18    \\
22      &   2006 May      15  & 871.43-871.44     &Mercator    &VRI  &   6     \\
23      &   2006 May      16  & 872.40-872.44     &Mercator    &VRI  &   45    \\
24      &   2006 May      17  & 873.40-873.43     &Mercator    &VRI  &   27    \\
25      &   2006 May      19  & 875.40-874.44     &Mercator    &VRI  &   33    \\
26      &   2006 May      20  & 876.41-876.43     &Mercator    &VRI  &   24    \\
27      &   2006 May      21  & 877.40-877.43     &Mercator    &VRI  &   30    \\
28      &   2006 May      23  & 879.40-879.41     &Mercator    &VRI  &   36    \\
\hline
\end{tabular}
\end{table}

\begin{table*}
\caption{Variable stars in M67 that fall inside CCD frames. See text for the
details.}\label{M67binary}
\begin{tabular}{lllllllllll}
\hline
Star Name  & S no.  &MMJ no. & Type          & Sp.T. & $\alpha_{2000}$   & $\delta_{2000}$   & m$_{\rm{V}}$ & B$-$V    & Period (days) \\
\hline
AH Cnc     &1282    &6027    & C             & F6.5V & 08 51 38   &   +11 50 57 &   13.33 &0.56  &   0.360           \\
EV Cnc     &1036    &5833    & SD            & F3V   & 08 51 28   &   +11 49 28 &   12.78 &0.49  &   0.441            \\
ES Cnc     &1082    &6493    & RS CVn        & F4V   & 08 51 21   &   +11 53 26 &   11.25 &0.42  &   1.068            \\
EX Cnc     &1284    &6504    & $\delta$ Sct  & A7V   & 08 51 34   &   +11 51 11 &   10.94 &0.22  &                      \\
EW Cnc     &1280    &5940    & $\delta$ Sct  &       & 08 51 33   &   +11 50 41 &   12.26 &0.26  &                      \\
           &1024    &5739    &               &       & 08 51 35   &   +11 50 32 &   12.72 &0.55  &   7.160            \\
           &1040    &6488    &               & G4III & 08 51 35   &   +11 50 32 &   11.52 &0.87  &   42.88            \\
           &1045    &5654    &               &       & 08 51 35   &   +11 50 32 &   12.54 &0.59  &   7.645            \\
           &1063    &5542    &               & G8IV  & 08 51 35   &   +11 50 32 &   13.79 &1.05  &   18.39              \\
           &1242    &5993    &               &       & 08 51 35   &   +11 50 32 &   12.72 &0.68  &   31.78            \\
           &1264    &5877    &               &       & 08 51 35   &   +11 50 32 &   12.09 &0.94  &   353.9              \\
\hline
\end{tabular}
\end{table*}

During the data reduction, we have studied all the nights and each
frame separately. The reduction and analysis of the frames have been
performed using the \textsc{IRAF} packages to subtract the
standard bias, dark, and divide flat--field, followed by aperture photometry.
10 comparison stars that were assured in the
previous studies were used. These comparison stars are
GSC\,814\,1205 (C$_1$), GSC\,814\,1425 (C$_2$), GSC\,814\,1311
(C$_3$), GSC\,814\,1146 (C$_4$), GSC\,814\,1981 (C$_5$),
GSC\,814\,1631 (C$_6$), GSC\,814\,1205 (C$_7$), GSC\,814\,1205
(C$_8$), GSC\,814\,1205 (C$_9$), and GSC\,814\,1205 (C$_{10}$).
For the Mercator observations only six of them were used.

\begin{figure}
\centering \includegraphics{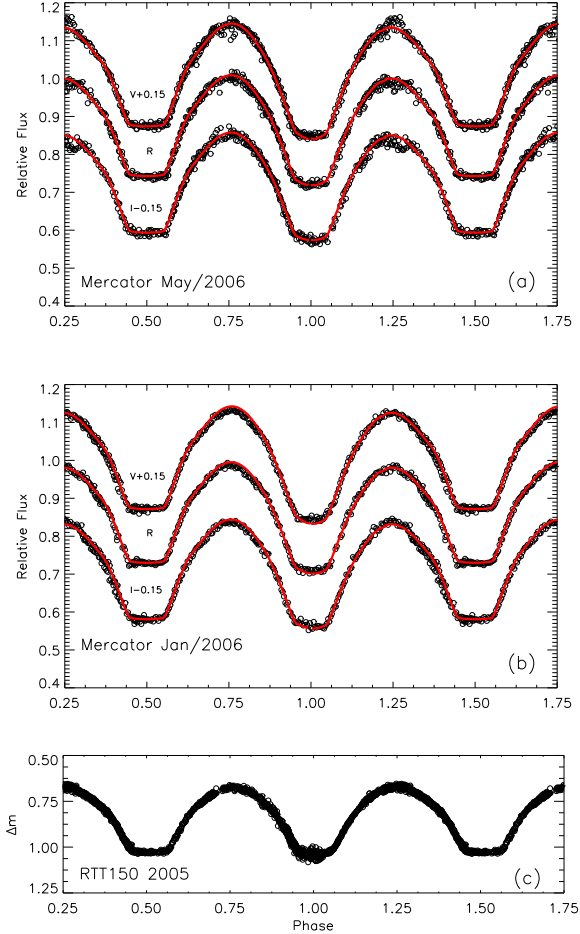} \vspace{-2.5cm}\caption{The observed and computed (solid line) light curves of AH~Cnc.
For the sake of comparison the light curves in V and I bands are shifted by a value of $+0.15$,
$-0.15$ in relative flux. }\label{m67_ahcnc_f1} \vspace{0.5cm}
\end{figure}

We observed the cluster on 28 nights, most of which successive. The data
from the successive nights revealed short time-scale light variations of AH
Cnc, EV Cnc, and ES Cnc. Variable stars falling inside the CCD frame are shown on the CM
diagram in Fig~\ref{11918fg1}. Some parameters of the variables are listed in
Table~\ref{M67binary}. V, B-V, and orbital periods given in the table
are obtained from Montgomery et al. (1993) and van den Berg et al.
(2002). Fig.~\ref{11918fg1} shows that AH Cnc and EV Cnc are at the turn-off
point of the cluster, whereas ES Cnc, EW Cnc, and EX Cnc lay in the
blue part of the diagram.
The light curves of AH Cnc obtained with both telescopes are shown in
Fig.~\ref{m67_ahcnc_f1} a-c.
The light curve of AH Cnc obtained by Whelan et al. (1979), using different telescopes and
photoelectric systems, does not show flat secondary minima as the light curve obtained in this
study. The difference in the light curves remains unexplained. The multi-colour observations of the total eclipse
light curves allow us to
determine accurate orbital parameters of the binary and the
physical parameters of the components. The orbital phases are calculated by using Eq.~(\ref{ahcnceq1}).

\begin{figure}
\centering \includegraphics{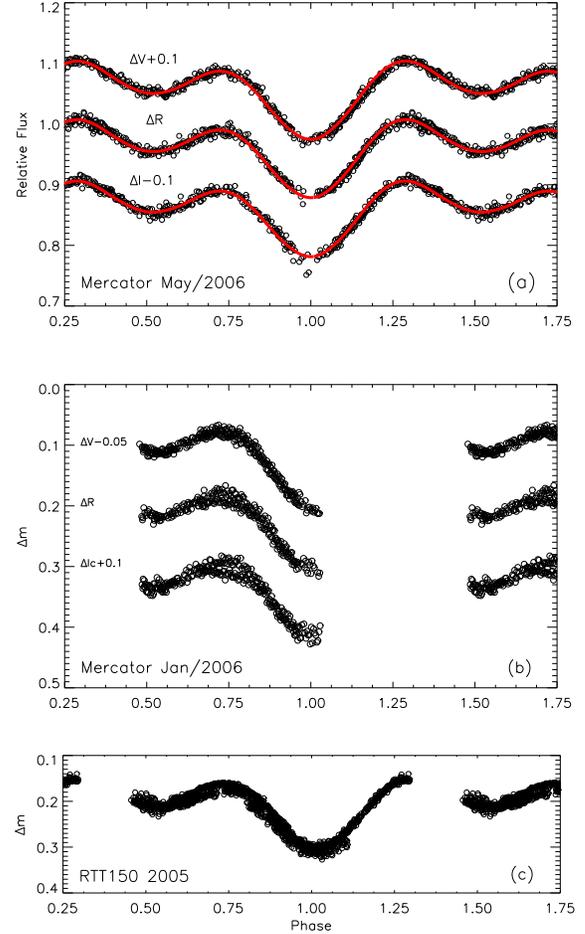} \vspace{-2.5cm}\caption{Same as Fig~\ref{m67_ahcnc_f1}, but for EV Cnc.
V and I bands are shifted by a value of +0.1 and -0.1 in relative flux.}\label{m67_evcnc_f1}
\end{figure}

\begin{figure}
\centering \includegraphics{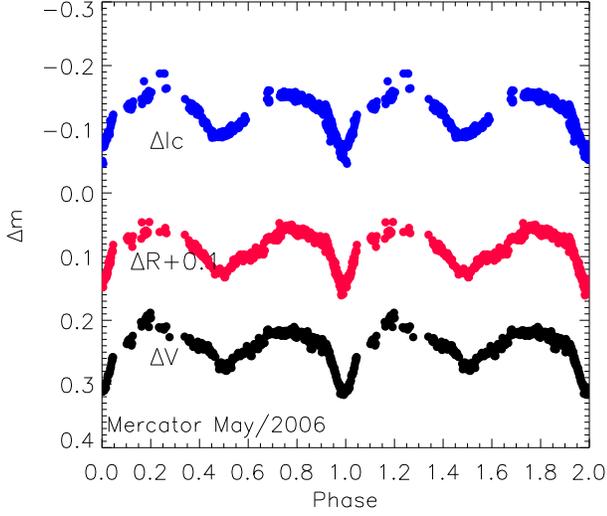} \caption{Same as Fig.~\ref{m67_ahcnc_f1} but for RS CVn type binary ES Cnc. See text for the details.}\label{11918fg4}
\end{figure}

The light variations of EV Cnc obtained with the Mercator telescope
and RTT150 are shown in Fig.~\ref{m67_evcnc_f1} The phases
are calculated using Eq.~(\ref{evcnceq1}).
The light curve obtained in May 2006 shows the full light variation
of the binary. Although the phase coverage of the Mercator observations made in January and that of the
RTT150 observations is not complete, these light curves are still useful. The observed difference in the
maxima levels are in agreement with
the observations given in the literature (see, van den Berg et al.
2002). We observe asymmetry during the ingress and egress of the secondary
minimum. We will discuss it again in the last section.
The difference in minima is conspicuous in the light curve of EV Cnc since the
system is of $\beta$ Lyr type rather than WUMa type.
The light variations of ES Cnc, obtained with the Mercator telescope
are shown in Fig.~\ref{11918fg4}. The phases are calculated using
the new ephemeris given by Eq.~(\ref{escnceq1}). The observations of the system made
in May (Fig.~\ref{11918fg4}) cover  the full phase, so
it gives the opportunity to study its light curve in detail.
The levels of max~I and max~II shown in Fig.~\ref{11918fg4}
are different. This is believed to be an activity indicator of the system.

\begin{figure}
\centering \includegraphics{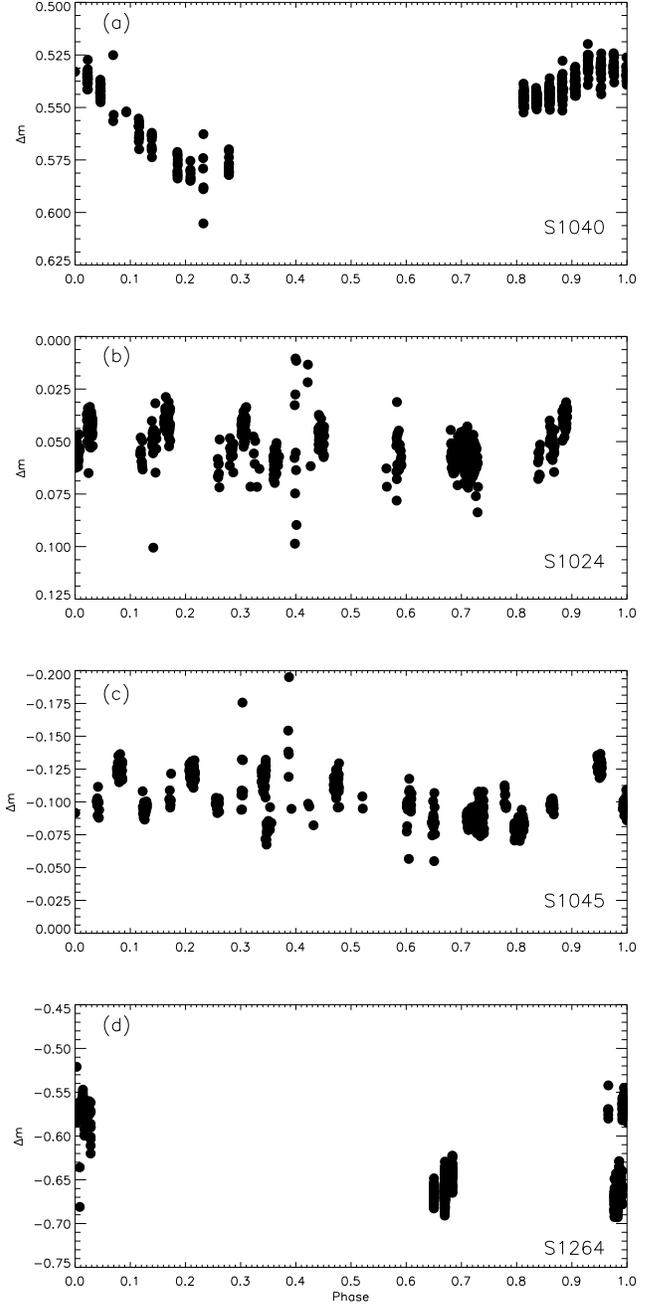} \caption{Observed V light curves of
(a) S1040, (b) S1024, (c) S1045, and (d) S1264.}\label{m67_s1024_f1}
\end{figure}

Long-period variables (S1024, S1040, S1045, S1063, S1242, and S1264)
that fall inside the CCD frame are given in Table~\ref{M67binary}.
The shortest period system is S1024 (7.16 days) and the
longest one with 354 days is S1264.
In the CM diagram of the open cluster, the systems
S1063 and S1113 lay well below the subgiant branch.  Mathieu et al.
(2003) studied these subgiant members of M67 spectroscopically and obtained
their orbital parameters. S1063 is a single-lined binary with
an orbital period of 18.4 days and an eccentricity 0.21.
The system S1113 is a double-lined binary and it has a
circular orbit with 2.8 days period. During the observations made
with the Mercator telescope the systems did not fall inside the CCD
and with the RTT150 the system
S1063 fell within the CCD frame. Their long periods require long term observations of them to retrieve information from their light
curves. The light variation of S1040 obtained in May, 2006
with the Mercator telescope is shown in Fig.~\ref{m67_s1024_f1}a.
The light variations of S1024, S1045, and S1264 are
shown in Fig.~\ref{m67_s1024_f1}b-d. The observed variations in these systems are not conspicuous
as in the case of S1040 (Fig.~\ref{m67_s1024_f1}a) though small amplitude variations
are still detectable. We do not have enough
data to analyze these long period variables, therefore,
new observations of them are important for evolutionary study of them.
\begin{table}
\caption{New times of the primary and secondary minima for AH Cnc and EV Cnc in
HJD-2\,400\,000.} \vspace{0.25cm} \label{m67tab2}
\begin{center}
\begin{tabular}{llllll}
\hline
& Times of minima  & error            &  Filter   \\
\hline
AH Cnc    &    &   &     &           \\
&53426.39314         &$\pm$ 0.00008    &    R$_c$       \\
&53489.29283         &$\pm$ 0.00008    &   R$_c$         \\
&53683.58142         &$\pm$ 0.00006    &   R$_c$        \\
&53750.62330         &$\pm$ 0.00025    &   I           \\
&53750.62366         &$\pm$ 0.00019    &   V            \\
&53750.62366         &$\pm$ 0.00014    &   R            \\
EV Cnc              &   &                   &              \\
&53425.4532         &$\pm$ 0.0003    &     R$_c$           \\
&53426.3392         &$\pm$ 0.0002    &     R$_c$            \\
&53682.6029         &$\pm$ 0.0003    &     R$_c$         \\
\hline
\end{tabular}
\end{center}
\end{table}
\section{Data analysis}
\subsection{New Ephemeris and Period change analysis of close binaries}
The orbital period of a binary system can vary because of mass transfer, mass loss, the presence of a third body etc. If the variations are large enough
they can be studied in O-C ($\Delta T = T_o-T_c$) diagrams.
The mass transfer between the components show a parabolic variation in the O-C diagram,
a third body in the system represent a sinusoidal variation.
Hence, a sinusoidal variation superimposed on an upward/downward parabola suggests mass transfer
between the components of a binary system that has a third body orbiting it (e.g. XY Leo (Yakut et al., 2003),
TY Boo (Yang et al., 2007), V839 Oph (Akalin \& Derman, 1997), BB Peg (Kalomeni et al., 2007)).
Usually, ephemeris of such a system is written as

\begin{equation}
MinI = T_o +P_oE + \frac{1}{2} \frac{dP}{dE}E^2 + \tau,
 \label{Eqper1} \end{equation}
where the first two terms are linear, the third one is due to the period increase/decrease and the last term $\tau$,
is the light-time term, which was formalized by Irwin (1952) as
\begin{equation}
\tau = \frac {a_{12} \sin
i'}{c}  \left[ \frac {1-{e'}^2}{1+{e'} \cos v'} \sin \left( v' +
\omega' \right) +{e'} \sin \omega' \right].
 \label{Eqper2}
\end{equation}

In Eq.~(\ref{Eqper1}),  $T_{\rm o}$ is the starting epoch for the primary minimum, $E$ is the integer
eclipse cycle number, $P_{\rm o}$ is the orbital period of the
eclipsing binary. In Eq.~(\ref{Eqper2}),  $a_{12}$, $i'$, $e'$, and $\omega'$
are the semi-major axis, inclination, eccentricity, and the
longitude of the periastron of the eclipsing pair around the third body,
and $v'$ is the true anomaly of the position of the center of
mass, time of periastron passage $T'$ and orbital period $P'$ are
the unknown parameters in Eq.(\ref{Eqper2}).

\subsubsection{The low-temperature contact binary AH Cnc (S1282)}
\begin{figure}
\centering
{\rotatebox{0}{\resizebox{8.5cm}{!}{\includegraphics{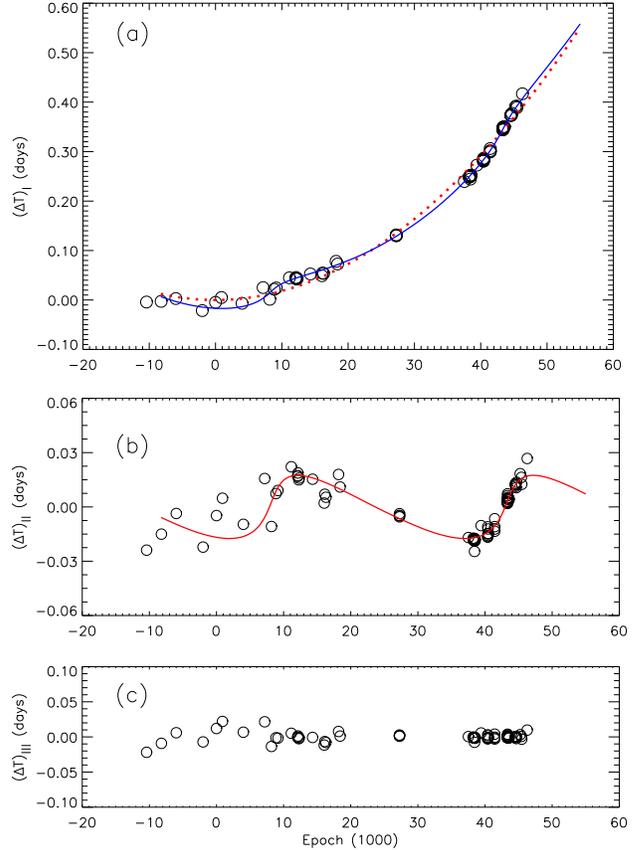}}}}\\[-5.5cm]
\caption{(a) The ($\triangle T)_{I}$ diagram of the times of
mid-eclipses for AH Cnc, (b) ($\triangle T)_{II}$ variation of the system after parabolic variation removed, and (c) final residual.} \label{11918fg6}
\end{figure}

In this study, we obtained six minima times for AH Cnc. They are
listed in Table~\ref{m67tab2} with their errors.
All the available times of light minimum of AH~Cnc were collected from the literature
 and are listed in Table~\ref{ahcnc_mintimes}. We improved the new linear ephemeris as:
\begin{equation}
\textrm{Min I} = \textrm{HJD} \,24 \, 53750\fd6239(7) +
0\fd3604583(2)\times E. \label{ahcnceq1}
 \end{equation}
In an earlier study, Qian et al. (2006) presented 50 minima times, while we collected 76 times of minima. We merged all these data to analyze the O-C variation. For the sake of comparison with Qian et al. analysis, differences between the observed
and calculated times ($\Delta T$) were estimated using the ephemeris given by Kriener et al. (2001).
They are plotted versus time in Fig.~\ref{11918fg6}.
The $(\Delta T)_I$ values show a parabola-like curve (Fig.~\ref{11918fg6})
which indicates the existence of mass transfer in the system.
The upward parabola shows that the mass transfer occurs from the less massive
component to the more massive one.

\begin{table}
\caption{Times of the primary and secondary minima in
JD$_{hel}$ (HJD- 2\,400\,000) for AH Cnc.}\vspace{0.25cm}
\label{ahcnc_mintimes}
\scriptsize
\begin{tabular}{llllll}
\hline
HJD         &meth &  Ref    &HJD         &meth &  Ref \\
\hline
33626.364	&	pg	&	1	&	51939.1411	&	ccd	&	12	\\
34421.307	&	pg	&	1	&	51940.2221	&	ccd	&	12	\\
35219.318	&	pg	&	1	&	51940.2278	&	ccd	&	12	\\
36656.352	&	pg	&	1	&	51956.9876	&	ccd	&	13	\\
37378.322	&	pg	&	2	&	51957.1642	&	ccd	&	13	\\
37699.300	&	pg 	&	2	&	51958.0662	&	ccd	&	13	\\
38820.7850	&	pe	&	3	&	51958.2465	&	ccd	&	13	\\
39964.300	&	pg	&	1	&	51959.1478	&	ccd	&	13	\\
40329.398	&	pg	&	1	&	52314.1944	&	ccd	&	12	\\
40570.9104	&	pe	&	4	&	52314.2001	&	ccd	&	13	\\
40678.3230	&	pe	&	5	&	52315.0943	&	ccd	&	12	\\
41396.332	&	pg	&	5	&	52315.2771	&	ccd	&	12	\\
41740.908	&	pg	&	6	&	52995.2842	&	ccd	&	13	\\
41742.8900	&	pe	&	6	&	52996.3640	&	ccd	&	13	\\
41752.8040	&	pe	&	6	&	52997.2642  &   ccd &   13  \\
41797.6750	&	pe	&	6	&	53001.2290	&	ccd	&	13	\\
41815.6980	&	pe	&	6	&	53004.1135	&	ccd	&	13	\\
42537.3000	&	pe	&	6	&	53004.2951	&	ccd	&	13	\\
43163.554	&	pg	&	5	&	53005.1948	&	ccd	&	13	\\
43192.214	&	pg	&	5	&	53005.3747	&	ccd	&	13	\\
43256.371	&	pg	&	5	&	53006.2758	&	ccd	&	13	\\
43931.492	&	pg	&	5	&	53007.1818	&	ccd	&	13	\\
44015.288	&	pg	&	5	&	53008.2609	&	ccd	&	13	\\
47200.6990	&	ccd	&	7	&	53009.3435	&	ccd	&	13	\\
47200.8790	&	ccd	&	7	&	53047.3702	&	ccd	&	14	\\
47203.7620	&	ccd	&	7	&	53083.0540	&	ccd	&	12	\\
47203.9440	&	ccd	&	7	&	53084.1348	&	ccd	&	12	\\
50904.288	&	pg	&	8	&	53426.3926	&	ccd	&	12	\\
51159.3059	&	ccd	&	9	&	53426.3931	&	ccd	&	14	\\
51177.1469	&	ccd	&	9	&	53437.7466	&	ccd	&	15	\\
51177.3275	&	ccd	&	9	&	53439.1876	&	ccd	&	12	\\
51179.3103	&	ccd	&	9	&	53442.7929	&	ccd	&	16	\\
51229.7727	&	ccd	&	10	&	53471.0869	&	ccd	&	12	\\
51231.7490	&	ccd	&	10	&	53489.2928	&	ccd	&	14	\\
51245.8122	&	ccd	&	10	&	53683.5814	&	ccd	&	14	\\
51250.6798	&	ccd	&	10	&	53750.6235	&	ccd	&	14	\\
51273.0268	&	ccd	&	9	&	53765.3982	&	ccd	&	17	\\
51585.1847	&	ccd	&	11	&	54060.9837	&	ccd	&	18	\\
\hline
\vspace{0.05cm}
\end{tabular}\\
References for Table ~\ref{ahcnc_mintimes}: 1-Kurochkin (1979), 2-Kurochkin (1965),
3-Eggen (1967), 4-Millis (1972), 5-Kurochkin (1970), 6-Whelan et al. (1979),
7-Estimated from light curves of Gilliland et al. (1991), 8-Diethelm (1998), 9-Youn et al. (2003),
10-Blake (2002), 11-Csizmadia et al. (2002), 12-Qian et al. (2006), 13-Zhang et al. (2005),
14-Krajci (2005), 15-Krajci (2006), 16-Nelson (2006), 17-Biro et al. (2006), 18-Nelson (2007).
\end{table}

Recently, the existence of a third body was reported by Qian et al. (2006) and Pribulla \& Rucinski (2006).
We used the linear ephemeris to construct the binary's O-C diagram. It shows
almost a sine-like variation superposed on an upward parabola. A
sine-like variation in the O-C curve, where both primary and secondary minima follow the same trend,
suggests a light time effect due to the presence of a tertiary component.

\begin{table}
\caption{Orbital elements of the tertiary component in AH Cnc.} \label{m67_ah_tab3b}
\begin{tabular}{llll}
\hline
Parameter            &Unit               & Value            &Error                        \\
\hline $T_o$         & [HJD]             & 2437378.3295     &$\pm0.0012$               \\
$P_o$                & [day]             & 0.3604360        &$\pm0.0000001$                \\
$P'$                 & [year]            & 34.7             &$\pm0.2$                      \\
$T'$                 & [HJD]             & 40374            &$\pm87$                   \\
$e'$                 &                   & 0.68             &$\pm0.03$                         \\
$\omega'$            & [$^\circ$]        & 9.9              &$\pm2.4$                   \\
$a_{12} \sin i'$     & [AU]              & 4.1              &$\pm0.1$             \\
$f(m)$               & [M$_{\odot}$]     & 0.056            &$\pm0.005$                      \\
$m_{3;i'=30^\circ}$  & [M$_{\odot}$]     & 1.57             &                 \\
$m_{3;i'=90^\circ}$  & [M$_{\odot}$]     & 0.60             &                 \\
$Q$                  & [c/d]             & $1.82\times 10^{-10}$&  $0.02\times 10^{-10}$               \\
\hline
\end{tabular}
\end{table}

The light elements from Eq.~(\ref{Eqper1}) and Eq.~(\ref{Eqper2}) were determined using the
differential correction method. We used these equations with the values
given in Table~\ref{ahcnc_mintimes} and a weighted least squares solution to
derive the parameters shown in Table~\ref{m67_ah_tab3b}. The
1$\sigma$ standard errors, in the last digit are given in
parentheses. We assigned a weight of
5 to the photoelectric (pe) and CCD observations,  1 to the photographic (pg) measurements.
The ($\triangle T_{I}$) variation of the system is plotted in Fig.~\ref{11918fg6}a. The
dashed line in Fig.~\ref{11918fg6}a shows the secular increase of the binary's
orbital period while the solid line is both the secular
increase and the light-time effect of the tertiary component.
A period increase ${dP}/{dt} = +3.69(6) \times 10^{-7}$ d yr$^{-1}$ is found.
The contribution of the light-time effect, $(\triangle
T)_{\mathrm{II}}$, to the total period variations of the system is shown in
Fig.~\ref{11918fg6}b.

\subsubsection{The near-contact binary EV Cnc (S1036)}
No period variation analysis of EV Cnc has been presented in literature.  We collected all available times of minimum light from the literature (see Table~\ref{evcnc_mintimes}) and added our new times of minima (Table~\ref{m67tab2}).
The linear ephemeris of the system were improved as:

\begin{equation}
\textrm{Min I} = \textrm{HJD} \,24 \, 53682\fd6039(7) +
0\fd441437(6)\times E. \label{evcnceq1}
 \end{equation}

\begin{table}
\caption{Times of the primary (I) and secondary (II) minima in
JD$_{hel}$ (HJD- 2\,400\,000) for EV Cnc.}\vspace{0.25cm}
\label{evcnc_mintimes}\footnotesize
\scriptsize
\begin{tabular}{lrrll}
\hline
HJD         &  Epoch    &$\Delta T_I$&$\Delta T_{II}$  &   Ref.\\
\hline
51229.7723	&	-2299	&	0.0175	&	0.0098	&	1	\\
51231.7441	&	-2294.5	&	0.0028	&	-0.0048	&	1	\\
51250.7368	&	-2251.5	&	0.0139	&	0.0065	&	1	\\
51603.2028	&	-1453	&	-0.0047	&	-0.0077	&	2	\\
51604.0883	&	-1451	&	-0.0020	&	-0.0051	&	2	\\
52244.6052	&	0	    &	-0.0050	&	-0.0050	&	3	\\
53108.2776	&	1956.5	&	0.0030	&	-0.0026	&	4	\\
53425.4532	&	2675	&	0.0087	&	-0.0018	&	5	\\
53426.3392	&	2677	&	0.0118	&	0.0013	&	5	\\
53443.3450	&	2715.5	&	0.0224	&	0.0117	&	6	\\
53682.6029	&	3257.5	&	0.0234	&	0.0079	&	5	\\
54144.1280	&	4303	&	0.0299	&	0.0029	&	7	\\
54147.2024	&	4310	&	0.0142	&	-0.0128	&	7	\\
54147.6584	&	4311	&	0.0288	&	0.0017	&	7	\\
54150.3054	&	4317	&	0.0272	&	0.0001	&	7	\\
54151.6301	&	4320	&	0.0276	&	0.0004	&	7	\\
54152.0721	&	4321	&	0.0282	&	0.0010	&	7	\\
54152.5075	&	4322	&	0.0221	&	-0.0051	&	7	\\
\hline
\vspace{0.05cm}
\end{tabular}\\
References for Table ~\ref{evcnc_mintimes}: 1- Blake (2002), 2-Csizmadia et al. (2002), 3-Csizmadia et al. (2006), 4-Krajci (2005), 5-This study, 6-H\"ubscher et al. (2005), 7-Pribulla et al. (2008).
\end{table}

\begin{figure}
\centering
{\rotatebox{0}{\resizebox{8.5cm}{!}{\includegraphics{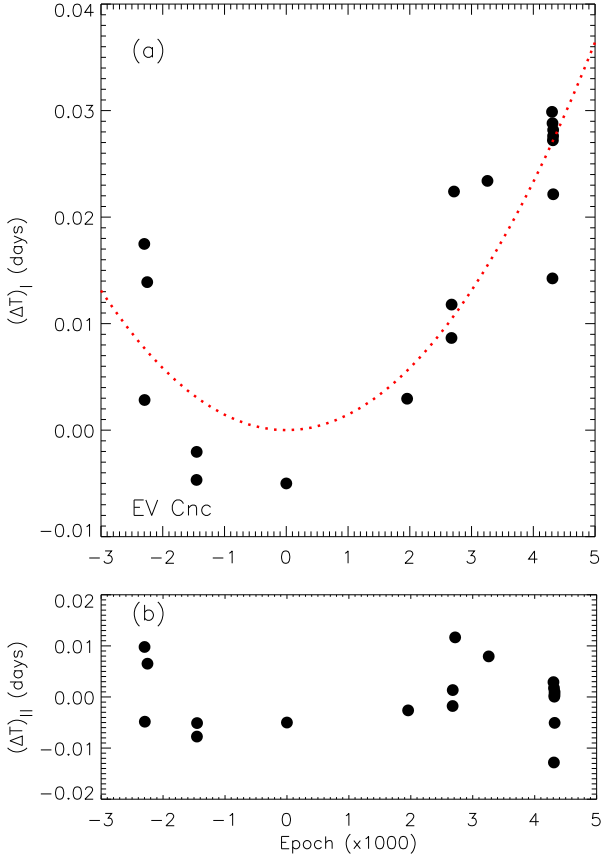}}}}\\[-5.0cm]
\caption{The ($\triangle T)_{I}$ diagram of the times of
mid-eclipses for EV Cnc.} \label{11918fg7}
\end{figure}

Since the system is a semi detached (or contact), continuing mass-transfer between
the components should affect the O--C variation. If the mass transfer
is large enough, a parabolic variation in O-C diagram can be detected. Fig.~\ref{11918fg7} indicates such a period increase over time. To obtain the light elements in Eq.~(\ref{Eqper1}) (first, second, and third term), the differential correction method was used. Applying this equation to the times of minima given in
Table~\ref{evcnc_mintimes} and using a least-squares solution
we obtained
\begin{equation}
\begin{array}{l}
\textrm{Min I} = 24\,52244.6102(6) + 0.4414340(2)\times E \\
+1.46(8)\times10^{-9}\times E^2.\label{EVCnc_Eq2}
\end{array}
\end{equation}

The variations of $\triangle T$ are displayed in Fig.~\ref{11918fg7}.
The observed O-C values in this figure are obtained with
linear elements. The dashed line in Fig.~\ref{11918fg7}a
represents the secular increase in the orbital period of the binary.
Fig.~\ref{11918fg7}b shows the residuals from the parabolic variation. Using Eq.~(\ref{EVCnc_Eq2})
a period change of ${dP}/{dt} = 2.4\times10^{-6}$ d yr$^{-1}$ is estimated. Such a value for the  period variation
is higher than expected
for semi-detached or contact binary systems. We need additional times of
minima than available to us now in order to determine the mass transfer
rate accurately.

\subsubsection{The blue straggler detached binary ES Cnc (S1082)}
We collected all the available minima times of ES Cnc and
listed them in Table~\ref{escnc_mintimes}. Using these minima times
and the first two terms in Eq.~(\ref{Eqper1}) we obtained the
new ephemeris of the system.

\begin{equation}
\textrm{Min I} = \textrm{HJD} \,24 \, 44643\fd2516(42) +
1\fd067797(1)\times E. \label{escnceq1}
 \end{equation}

\begin{table}
\caption{Times of the primary (I) and secondary (II) minima in
JD$_{hel}$ (HJD- 2\,400\,000) for ES Cnc.}\vspace{0.25cm}
\label{escnc_mintimes}\footnotesize
\scriptsize
\begin{tabular}{lrrll}
\hline
HJD         &  Epoch    &$\Delta T_I$ &   $\Delta T_{II}$ & Ref.\\
\hline
43191.036	&	-1360	&	0.0472	&	-0.0118	&	1	\\
44643.250	&	0	&	0.0478	&	-0.0016	&	1	\\
45325.586	&	639	&	0.0570	&	0.0122	&	1	\\
46773.492	&	1995	&	0.0208	&	-0.0144	&	1	\\
47861.605	&	3014	&	0.0415	&	0.0136	&	1	\\
47920.333	&	3069	&	0.0403	&	0.0127	&	1	\\
47944.869	&	3092	&	0.0168	&	-0.0106	&	1	\\
51218.744	&	6158	&	0.0048	&	-0.0009	&	1	\\
51539.078	&	6458	&	-0.0024	&	-0.0059	&	1	\\
51602.0961	&	6517	&	0.0152	&	0.0122	&	2	\\
51603.1655	&	6518	&	0.0168	&	0.0138	&	2	\\
53765.4321	&	8543	&	-0.0197	&	-0.0083	&	3	\\
54165.8639	&	8918	&	-0.0144	&	-0.0004	&	4	\\
54167.9957	&	8920	&	-0.0182	&	-0.0042	&	4	\\
54170.1339	&	8922	&	-0.0156	&	-0.0016	&	4	\\
54171.2019	&	8923	&	-0.0154	&	-0.0014	&	4	\\
54172.2696	&	8924	&	-0.0155	&	-0.0015	&	4	\\
54173.3370	&	8925	&	-0.0159	&	-0.0019	&	4	\\
\hline
\vspace{0.05cm}
\end{tabular}\\
References for Table ~\ref{escnc_mintimes}: 1- van den Berg et al.\ (2001), 2-Csizmadia et al. (2002), 3-Biro et al. (2006), 4-Pribulla et al. (2008).
\end{table}

\begin{figure}
\centering
\centering
\includegraphics{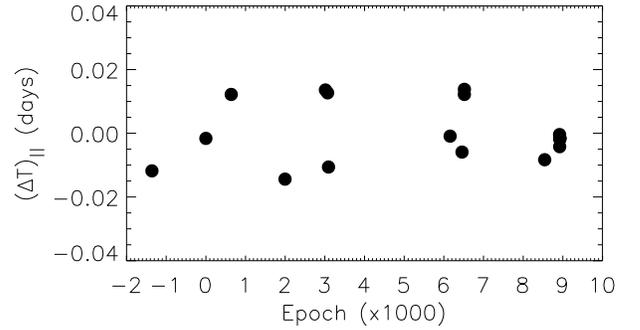}
\caption{The ($\triangle T)_{II}$ diagram of the times of
mid-eclipses for ES Cnc.} \label{m67_escnc_OC}
\end{figure}
Unlike AH Cnc and EV Cnc, ES Cnc is a detached binary system with no mass flow between the components.
However, it may lose  mass via a stellar wind. The mass loss from the system can
affect the O-C diagram, but this non-conservative effect is less effective than the effect of conservative mass transfer. The O-C variation of the system, after the corrections of
T$_{\rm o}$ and $P$, is shown in Fig.~\ref{m67_escnc_OC}.  Spectral studies of
the system show the presence of a third body with a 3.3\,yr period.  To
model the light-time effect, we need more times of minima than those we obtained
on successive nights.

\subsection{Light curve modeling of close binaries}
\subsubsection{AH Cnc}

In contrast to previous studies, we obtained a 28 day coverage of the
high-quality data, including four colour light curves, of the overcontact binary
system AH Cnc.  The multi-colour solution of the system allows us to determine
more accurate orbital and physical parameters of the binary than the earlier
studies. Fig.~\ref{m67_ahcnc_f1} shows that the system has a total
eclipse, which is an important tool to determine the mass ratio accurately
in the absence of a
good quality spectrum.

The light curve analysis of the system  has been made
by using the WD code in Mode~3 (Wilson \& Devinney 1971; Wilson
1994). The adopted values are: $T_\mathrm{1} = 6300$\,K
from the $B-V$ colour index, gravity darkening
coefficients, and albedos were chosen as $g_\mathrm{1}$ =
$g_\mathrm{2}$ = 0.32 (Lucy 1967) and $A_\mathrm{1}$ =
$A_\mathrm{2}$ = 0.5 (Rucinski 1969) and the logarithmic limb-darkening
coefficients ($x_\mathrm{1}$, $x_\mathrm{2}$) were obtained from
van Hamme (1993). The inclination $i$, temperature of the secondary
component $T_\mathrm{2}$, luminosities of the primary
component $L_\mathrm{1}$ (V, R, and I), the potential of the common
surface $\Omega$, and mass ratio were adjustable parameters.
V, R, and I light curves were analysed simultaneously. The results obtained after several
dozen iterations are given in Table~\ref{tab4}.

We first modelled the May LC and deduced the parameters.
Then we modelled the Jan LC of the system assuming the same orbital parameters and adding free spot and third
light parameters. The computed light curves (solid lines) with the parameters given in the last column of
Table~\ref{tab4} were compared with all observed light curves shown in Fig.~\ref{m67_ahcnc_f1}.
During the solution we consider the deep minimum as the one of the primary.
The synthetic light curves were created with the LC program.
The filling factor, f = 0.56, is given by
$(\Omega _{in} -\Omega)/ (\Omega _{in} -\Omega _{out} )$
and varies from 0 to unity from the inner to the outer critical surface.
This solution indicates that the system has a deep degree of contact.

\begin{table*}
\caption{Photometric elements of the over contact binary AH Cnc. See text for the details.}\label{tab4}
\begin{tabular}{lllllllllll}
\hline
Parameter                                             & W       & SS      & Y       &ZZD             & YA        & Q1       & Q2    &  P   & PS1  & PS2   \\
\hline
Geometric parameters:&&&&&&&&&&\\
$i$ ${({^\circ})}$                                    & 65      &86       &86       &82.8            &87.6       &90.0      &90.0   & 89.5    & 88.2 	        & 88.2(7) \\
$\Omega _{1,2}$                                       & -       &-        &2.078    &2.037           &2.008      &2.130     &2.088  & 2.023   & 2.093 	        & 2.093(3)  \\
$q$                                                   & 0.75    &0.157    &0.159    &0.149           &0.144      &0.160     &0.168  & 0.130   & 0.168 	        & 0.168(2) \\
$f$ (\%)                                              & 85      &70       &49       &65              &87         &37	    &59     &         & 56 	            & 56  \\
Radiative parameters:&&&&&&&&&&\\
$T_1$ (K)                                             &         &         &6300     &                &6300       &6300      &6300   & 6300    & 6300	        & 6300  \\
$T_2$ (K)                                             &         &         &6100     &6354            &6525       &6179      &6265   & 6368    & 6275 	        & 6275(90)	\\
Albedo$^*$ ($A_1=A_2$)                                &         &         &0.5      &                &0.5        &0.5       &0.5    & 0.5     &0.5 	            & 0.5\\
Gravity brightening$^*$ ($g_1=g_2$)                   &         &         &0.32     &                &0.32       &0.32      &0.32   & 0.25    &0.32	            & 0.32\\
$(\frac{l_1}{l_{\rm total}})$ :&&&&&&&&&&\\
~~~~~~~~~~~~~~~~~~$_V$                                &         &         &0.850    &0.830           &           &	 0.844  & 0.819	&         & 0.790 	        & 0.783     \\
~~~~~~~~~~~~~~~~~~$_R$                                &         &         &         &                &0.827      &	        &	    &         & 0.792 	        & 0.780      \\
~~~~~~~~~~~~~~~~~~$_I$                                &         &         &         &                &           &	        &	    &         & 0.795 	        & 0.783    \\
$(\frac{l_3}{l_{\rm total}})$ :&&&&&&&&&&\\
~~~~~~~~~~~~~~~~~~$_V$                                &         &         &         &                &           &	        &0.007  &         & 0.034      	    & 0.036     \\
~~~~~~~~~~~~~~~~~~$_R$                                &         &         &         &                &           &	        &	    &         & 0.031     	    & 0.040      \\
~~~~~~~~~~~~~~~~~~$_I$                                &         &         &         &                &           &	        &	    &         & 0.026     	    & 0.035    \\
Fractional radii             &&&&&&&&&&\\
$\bar{r}_{1}$                                         &         &         & 0.561   &0.572           &0.586      &0.5555    &0.5603 &         & 0.5584          & 0.5584(14)      \\
$\bar{r}_{2}$                                         &         &         &0.258    &0.260           &0.271      &0.2505    &0.2658 &         & 0.2531          & 0.2630(60)     \\
Spot parameters: &&&&&&&&&&\\
Colatitude   ${({^\circ})}$                           &         &         &         &                &           &	        &	    &         & 113(5)	        & 98(6)	    \\
Longitude    ${({^\circ})}$                           &         &         &         &                &           &	        &	    &         & 121(7)	        & 105(9)	       \\
Spot radius  ${({^\circ})}$                           &         &         &         &                &           &	        &	    &         & 35(2)	        & 24(3)		   \\
Spot temperature ($T_{spot}/T_{star}$)                &         &         &         &                &           &	        &	    &         & 0.91(1)	        & 0.92(1)		   \\
rms&                                                     &          &         &         &               &           &           &    & 0.018      &    0.031           \\
\hline
\end{tabular}
\end{table*}

The light curve of the system was modelled (see Table~\ref{tab4}) by Whelan et al. (1979), Sandquist \& Shetrone (2003), Youn et al. (2003), Zhang,
Zhang \& Deng (2005), Qian et al. (2006), Yakut \& Aerts (2006), and recently by Pribula et al. (2008) using one colour MOST space photometry.
In Table~\ref{tab4} they are shown as W, SS, Y, ZZD, YA, Q1- Qian et al. (2006) solution without third body, Q2- Qian et al. solution with third body, P, PS1- present solution (January) with third light, and PS2-present solution (May) with third light, respectively.
In the present study, three colour solutions are made and lead to more accurate parameters.

\begin{figure}
\centering \includegraphics{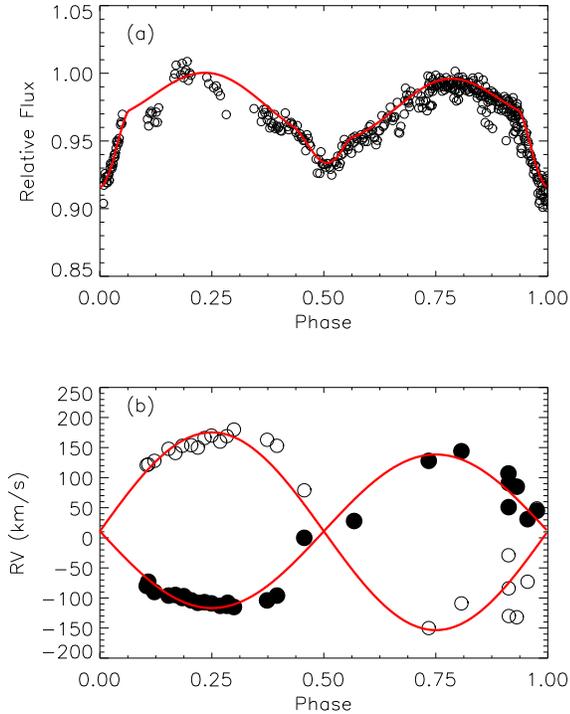} \caption{The observed V band and
computed light (a) and radial velocity curves (b) of ES Cnc . }\label{WDES} \vspace{0.5cm}
\end{figure}
\subsubsection{EV Cnc}
The light curve of EV Cnc is quite different from those of
normal contact or near--contact systems. Therefore, different
possibilities were taken into consideration. The minima are quite different, which may be an
indication of a semi-detached configuration rather than a contact one, as
proposed in the literature. The O'Connell effect is apparent in the
light curves of EV Cnc presented in the literature. This effect can
be explained by a mass transfer in a semi-detached system that
gives rise to a hot region, rather than a spot activity on
the surface of low-temperature contact binaries.

The first eleven cosine Fourier coefficients of our light curves of EV Cnc were calculated.
Rucincki (1993) and (Selam, 2004) have plotted the values of $a_4$ versus  $a_2$ for different contact degrees.
In their figure the system  EV Cnc lies outside of the contact region. In addition, we have tested
our $a_2$ and $a_4$ values in the Paczynski et al. (2006) diagram, where they plotted about 11000 close binaries.
The system is situated close to the semi-detached binaries rather than the contacts.
We used Mode 3, Mode 4, and Mode 5 of the WD code (see Table~\ref{tab5}) and we tested our results using Phoebe (Pr{\v s}a \& Zwitter, 2005).
Mode~5 in the DC code was used for the final light curve modeling.
This mode solves the light curves of semi-detached binaries,
where the secondary component fills its corresponding Roche lobe while the primary is detached.
The following parameters were fixed for the light curve analysis: the mean temperature of the hotter component is $T_h$ = 6900~K according to the colour index and the logarithmic limb darkening coefficients. The adjustable
parameters are the inclination, the mass ratio, the temperature of the
secondary component, the luminosities, the surface potential,
the values of the gravity darkening coefficients $g_1$ and $g_2$, and the values of the albedos.

The results of a simultaneous three colour light curve solution are summarized in Table~\ref{tab5}.
The results of different model
solutions (M) have been denoted with different letters. We have assigned Model A for the contact configuration,
Model B for the non-thermal contact, and finally Model C for the semi-detached configuration. Our analysis shows
that the configuration of EV Cnc is almost contact.
The model results (Model C, solid line) and observations are compared in Fig.~\ref{m67_evcnc_f1}a.
The asymmetries in the light curves indicate that the components may have hot or cold activity regions on their surfaces.
Hence, throughout the solutions the spotted model assumption was considered.
A hot spot and two cold ones determined on the surfaces of the cooler component.
The spot parameters given in Table~\ref{tab5} are common for all the three models.

\begin{table}
\caption{Photometric elements of EV Cnc. See text for the details.}\label{tab5}
\scriptsize
\begin{tabular}{llll}
\hline
Parameter                                             & Model A    & Model B & Model C   \\
\hline
$i$ ${({^\circ})}$                                    & 46.7        &48.6       &41.8(5)         \\
$\Omega _{1}$                                         & 3.30        & 3.133     & 2.709       \\
$q$                                                   & 0.69        &0.59       & 0.40(1)       \\
$T_{\rm h}$ (K)                                       & 6900        &6900       & 6900          \\
$T_{\rm c}$ (K)                                       & 5630        &5830       & 5200 (165)    \\
$(\frac{L_c}{L_c +L_h})_V$                            & 0.92        &           &  0.382       \\
$(\frac{L_c}{L_c +L_h})_R$                            & 0.89        &           &  0.443       \\
$(\frac{L_c}{L_c +L_h})_I$                            & 0.86        &           &  0.484       \\
$\bar{r}_{c}$                                         & 0.40        & 0.38      &  0.452(2)         \\
$\bar{r}_{h}$                                         & 0.33        & 0.29      &  0.300(3)          \\
Spot parameters             : &&       &        \\
three spots on secondary: &&       &        \\
Colatitude   ${({^\circ})}$                           &     &&49(3),119(6),87(4)                   \\
Longitude    ${({^\circ})}$                           &     &&203(10),177(12),11(3)                    \\
Spot radius  ${({^\circ})}$                           &     &&21(3),28(4),16(2)                      \\
Spot temperature $T_{spot}/T_{star}$                  &     &&0.89(3), 0.88(2),1.15(2)              \\
rms                                    & &&0.012\\
\hline
\end{tabular}
\end{table}

\subsubsection{ES Cnc}
ES Cnc is an RS CVn type active close binary.
The system is brighter than AH Cnc and EV Cnc; therefore,
RTT150 observations were saturated.
The radial velocity curve of the system has been obtained by van den Berg et al. (2001)
and Sandquist et al. (2003).
Full light curves of the system were obtained with the Mercator telescope.
The light variations of the system show small variations on a time scale of weeks.

We have calculated the Fourier coefficients of the light curve.
It falls in the detached region on the $a_4$ vs $a_2$ plane. The light curve of ES Cnc has been well
analyzed before by van den Berg et al. (2001) using the Eclipsing Light Curve
code of Orosz \& Hauschildt (2000), by Sandquist et al. (2003) with Nightfall and by Pribulla et al. (2008).
All researchers solved the system assuming the third light effect and
stellar spots. We used the WD code under the assumptions of the presence of a third light and spots.
The adopted values are: $T_\mathrm{1} = 7325$\,K, gravity darkening
coefficients, and albedos were chosen as $g_\mathrm{1}$ =1,
$g_\mathrm{2}$ = 0.32  and $A_\mathrm{1}$ =1, $A_\mathrm{2}$ = 0.5 and the logarithmic limb-darkening
coefficients ($x_\mathrm{1}$, $x_\mathrm{2}$) were obtained from
van Hamme (1993). The semi major axis of the relative orbit, $a$; binary center-of-mass radial
velocity, $V_\gamma$;  inclination, $i$; temperature of the secondary
component, $T_\mathrm{2}$; luminosities of the primary
component, $L_\mathrm{1}$; the potentials of the components, $\Omega_1$ and $\Omega_2$,
and spot parameters (latitude, longitude, size, and temperature factor) were adjustable parameters.
The results of a simultaneous solution of LC and RV are given in Table~\ref{tab-ESCNC}.
The comparison of the observations and model are shown in Fig.~\ref{WDES}.

\begin{table*}
\caption{Photometric parameters of detached system ES Cnc.}\label{tab-ESCNC}
\begin{tabular}{lllll}
\hline
Parameter                      & van den Berg et al. (2001)     &Sandquist et al. (2003) & Pribulla et al. (2008)  & Present study \\
\hline
$i$ ${({^\circ})}$                                   &64(1)         &68(1)       &66.2  & 68.3(2.8)       \\
$\Omega _{1}$                                        &-             &-           &3.532 & 4.12(21)      \\
$\Omega _{2}$                                        &-             &-           &3.922 & 4.67(36)      \\
$a$ ($\textrm{R}{_\odot}$)                                                   &7.2(4)        &-           &      & 6.629(144)      \\
$q$                                                  &0.63(4)       &-           &0.68  & 0.777(16)      \\
$V_\gamma$                                          &0.63(4)       &-           &      & 11(1)      \\
$T_{Aa}$ (K)                                         &6480(25)      &7325(50)    &7325  & 7325(50)      \\
$T_{Ab}$ (K)                                         &5450(40)      &6000(200)   &5543  & 5975(200)      \\
$(\frac{l_1}{l_1 +l_2})_V$                           &-             &-           &      & 0.82         \\
$(\frac{l_3}{l_{total}})_V$                          &-             &-           &0.68  & 0.42(15)         \\
$\bar{r}_{1}$                                        &-             &-           &      & 0.309(21)       \\
$\bar{r}_{2}$                                        &-             &-           &      & 0.222(18)        \\
Spot parameters (spots on cool star): &\\
Colatitude   ${({^\circ})}$                           &             &            &             &92,135             \\
Longitude    ${({^\circ})}$                           &             &            &179.9, 296.6 &328, 57                \\
Spot radius  ${({^\circ})}$                           &             &            &11.5, 12.5    &33, 30                  \\
Spot temperature ($T_{spot}/T_{star}$)                &             &            &0.80, 0.80    &0.83, 0.83                  \\
rms                                       &       &            &             &          0.175                    \\
\hline
\end{tabular}
\end{table*}

\subsection{Frequency analysis of the $\delta$ Scuti stars EX Cnc (S1284) and EW Cnc (S1280)}

For both stars, we carried out a frequency analysis and a mode identification based on the multi-colour data that are available.
Bruntt et al. (2007), hereafter referred to as B07, have already conducted a successful multi-site campaign on EX Cnc and EW Cnc. Their data provided 26 frequencies for EX Cnc and 41 for EW Cnc, but they could not carry out a photometric mode identification due to the fact that their measurements were acquired in only one filter. Therefore they were not able to constrain fundamental parameters of the stars and to provide a unique seismic model.

Our data consist of multi-colour data taken simultaneously in the $V$, $R$, and $I$ filters at the Mercator telescope. In contrast to B07, our observations are single-site data which complicates the search for pulsation frequencies due to daily aliasing problems. Our aim was to use the frequency values that were determined by B07 and to carry out a mode identification using the method of amplitude ratios and phase differences (Dupret et al., 2003; Daszynska-Daszkiewicz et al., 2003).

To determine differential magnitudes of the stars in the different filters, we had six different comparison stars on the CCD-frames at our disposal. Most of these comparisons proved to show systematic trends on the time scales of days. These trends are smallest for the comparisons C3 and C5, which we therefore used to determine differential magnitudes. For our search for pulsation frequencies we used Fourier techniques for non-equidistant measurements (discrete Fourier transformation) and multi-periodic least-squares fitting techniques for sinusoidal variations. The frequency search was carried out in the following iterative manner. We selected the highest frequency peak in the Fourier spectrum, compute a least-squares fit of this peak and all previously found frequencies by improving frequencies, amplitudes and phases of a sum of sinusoids. In case of aliasing, we selected the frequency value that resulted in the lowest residuals after pre-whitening. We then pre-whiten the original data with the determined least-squares fit and computed a Fourier spectrum of the residuals. This procedure was carried out iteratively until the amplitude signal-to-noise ratio (SNR) of the highest frequency peak remained below a value of 4.0 (Breger 1999). We used the software package FAMIAS (Zima 2008) for the frequency search as well as for the mode identification.
\subsubsection{EX Cnc}
We used the differential light curve $EX-C5$ in $V$ of EX Cnc and C5 for our frequency analysis since these data showed the best long term stability. The highest peak is at $f_1=20.61$~c/d which corresponds to the frequency $g_2=20.63$~c/d detected by B07. The highest peak detected by B07 is at $g_1=19.61$~c/d, which is exactly 1~c/d separated from $f_1$. Due to the strong daily aliasing in our data, we cannot exclude that 19.6~c/d is the correct frequency value. We tested both scenarios and preferred the frequency solution with $20.61$ c/d due to the lower residuals after pre-whitening with the least-squares fit. The residuals show no significant frequency peaks. We also searched in the residuals for frequencies with a SNR$<4$ that were reported by B07, but could not find any acceptable agreement. The amplitudes, phases, and the SNR of $f_1$ in the different filters are displayed in Table~\ref{tab:excncfrequencies}. The phase is indicated in units of $2\pi$ and computed relative to the zero point at HJD 2450000. The uncertainty of the measurements in units of the last digit is indicated in brackets. The noise for the SNR calculation was computed from the mean of the amplitude of the pre-whitened Fourier spectrum in a 3~c/d box around each frequency.

\begin{table}
\caption{Results of the frequency analysis for EX Cnc. See text for the details.}\label{tab:excncfrequencies}
\begin{tabular}{lllll}
\hline
Frequency (c/d)  & filter &  $A$ (mmag)   & $\phi$ & S/N \\
\hline
               &  $V$    & 5.6(2)    & 0.408(8)    & 6.5      \\
f$_1=$20.605(11) &  $R$    & 4.2(3)    & 0.404(11)   & 4.7      \\
               &  $I$    & 4.0(3)    & 0.411(12)   & 6.5      \\
\hline
\end{tabular}
\end{table}

The main criterion for the mode identification of $\delta$~Sct stars is the phase difference in different photometric passbands. The phase difference can be used to distinguish between different values of the degree $\ell$. As can be seen in Table~\ref{tab:excncfrequencies}, the uncertainty of the phase values is of the same order as the difference between the three filters. We therefore conclude that the quality of the data is insufficient to acquire a conclusive mode identification for this frequency.

\subsubsection{EW Cnc}

We chose the comparison star C3 for this object due to the best long term stability of the differential light curve.
The two highest peaks in our data at $f_1=18.82$ and $f_2=20.19$~c/d and with amplitudes of about 3 and 2 mmag, respectively, in V agree well with the frequencies of highest amplitude detected by B07. We find an additional frequency at $f_3=37.60$~c/d, whereas B07 report two frequencies very closeby, at 37.617 and 37.654. Although, the theoretical frequency resolution of our data is 0.01~c/d (Loumos \& Deeming, 1978) - enough to separate the two frequencies, no frequency peak remains close to these values after pre-whitening with $f_3$. We find two more frequencies that were reported by B07, namely at $f_4=31.43$ and $f_5=27.86$~c/d. After pre-whitening with the five frequencies, a significant peak that was not found by B07 with a SNR of 6.5 at $f_6=37.13$~c/d remains. The residuals after pre-whitening show no significant peaks and no additional frequencies can be identified with frequencies reported by B07. The Cousins $R$ and $I$ light curves are much noisier than the $V$ data. Therefore, much less frequencies could be detected in these filters. We relaxed the SNR criterion for significance to $SNR \ge 3.5$ for the frequencies that were already detected in the $V$-band. Table~\ref{tab:ewcncfrequencies} reports the result of the frequency search for the light curves of EW Cen.

\begin{table}
\caption{Same as Table~\ref{tab:excncfrequencies} but, for EW Cnc.}\label{tab:ewcncfrequencies}
\begin{tabular}{lllll}
\hline
Frequency (c/d)  & filter &  $A$ (mmag)   & $\phi$ & S/N \\
\hline
               &  $V$    & 3.0(3)    & 0.912(15)   & 7.1      \\
f$_1=$18.824(11) &  $R$    & 1.7(4)    & 0.881(31)   & 3.5      \\
               &  $I$    & 1.4(4)    & 0.928(45)   & 3.5      \\
\hline
               &  $V$    & 2.1(3)    & 0.097(21)   & 4.9      \\
f$_2=$20.196(11) &  $R$    & 1.5(4)    & 0.072(35)   & 3.5      \\
               &  $I$    & 1.6(4)    & 0.151(38)   & 3.9      \\
\hline
f$_3=$37.606(11) &  $V$    & 1.6(3)    & 0.237(28)   & 7.6      \\
               &  $I$    & 2.2(4)    & 0.217(27)   & 3.8      \\
\hline
f$_4=$31.438(11) &  $V$    & 1.6(3)    & 0.047(29)   & 7.0      \\
               &  $R$    & 1.6(4)    & 0.967(34)   & 3.7      \\
\hline
f$_5=$27.862(11) &  $V$    & 1.2(3)    & 0.010(40)   & 3.9      \\
\hline
f$_6=$37.135(11) &  $V$    & 1.4(3)    & 0.793(32)   & 6.6      \\
\hline
\end{tabular}
\end{table}

Also for the derived pulsation phases of EW Cnc, the formal statistical uncertainties that were derived from the least-squares fitting, are of the same order as the phase difference between the different filters. We therefore, unfortunately, have to conclude that also for this star the quality and amount of data are not sufficient for a successful photometric mode identification.

\section{Results and Conclusions}

We presented 28 days long term multi-colour observations for the solar age galactic open cluster
M67.

\begin{table*}
\caption{Astrophysical parameters for the close binaries AH Cnc and ES Cnc.}\label{pyspar}
\begin{tabular}{llllll}
\hline                                  &                 & ~~~~~~~~~~~~~AH Cnc  &          	 &  ~~~~~~~~~~~~~ES Cnc &		  \\
Quantity &Unit&Pri. & Sec.  &Pri. & Sec. \\
\hline
Mass   (M)                               &$\rm{M_{\odot}}$      & $1.47\,(15)$      & $0.25\,(3)$    & $1.94\,(13)$   & $1.50\,(9)$    \\
Radius  (R)                              &$\rm{R_{\odot}}$      & $1.40\,(9)$      & $0.68\,(5)$    & $2.05\,(7)$   & $1.47\,(6)$  	 \\
Effective Temperature (T$_e$)            &$\rm{K}$              & $6300$           & $6275\,(90)$    & $7325\, (50)$  & $5975\,(200)$    \\
Luminosity     (L)                       &$\rm{L_{\odot}} $     & $2.78\,(50)$      & $0.64\,(11)$     & $10.8\,(7)$  & $2.5\,(6)$ 	   \\
Surface Gravity ($log\,g$)               &cgs                   & 4.31             & 4.17            & 4.10 	       & 4.28  	        \\
Bolometric Magnitude (M$_{bol}$)         &mag                   & 3.64             &5.23             & 2.16            & 3.77 	   \\
Visual Magnitude       (V)               &mag                   & 13.60            &15.19            & 12.05           & 13.80 	   \\
Absolute Magnitude     (M$_{V}$)         &mag                   & 3.80             & 5.39            & 2.25            & 3.95 	   \\
Bolometric Correction (B.C.)             &mag                   &-0.16             & -0.16           & -0.09           & -0.18 	   \\
Period Change Rate ($\dot{P}$)          &d\,yr$^{-1}$          & ~~~~~~~$+3.7 \times 10^{-7}$          &       & ~~~~~~~ $+5.1 \times 10^{-10}$     &  \\
Mass Transfer Rate ($\dot{M}$)          &$\rm{M_{\odot}}$ yr$^{-1}$ & ~~~~~~~$+9.4 \times 10^{-8}$      &     &    &    \\
Distance (d)                             &pc                     &                &                        & ~~~~~~~~~~~~857(33)             &   	 \\
\hline
\end{tabular}
\end{table*}

Radial velocity studies of EV Cnc and AH Cnc made by Blake (2002) and Whelan et al. (1979), respectively.
Because of their faintness and configuration,
the parameters were not as accurate as those of ES Cnc.
It is useful, therefore, to deduce the mass ratio from a LC analysis.
AH Cnc is an overcontact binary showing total eclipses.
A multi-colour analysis was made.
The inclination angle of the system was determined ${88^\circ}$ and the mass ratio of the components was estimated 0.17.
We have collected 70 minima times from literature and did a period analysis.
The O-C diagram of AH Cnc show shallow sine-like variation superimposed
on a parabolic variation. The mass transfer rate is estimated to be ${dM}/{dt}=+9.4 \times 10^{-8}$ $M_{\odot}$yr$^{-1}$.
Using the brightness of the systems (Table~\ref{M67binary}) their distance modules, and bolometric corrections we deduced
the absolute physical parameters of the components of AH Cnc (Table~\ref{pyspar}).
For the solar values, we have taken $T_{\rm e}$ = 5780~K, $M_{\rm bol}$ = $4^{\rm m}.75$, and we adopted E(B-V)=41 mmag for M67 (Taylor, 2007).
The mass of the primary star of AH Cnc has been estimated from the mass-radius relation of
well-known LTCBs (Yakut, 2005, 2006) and we obtained a value of 1.33 $M_{\odot}$.
Zhang et al. (2005), using the distance modulus, reported a mass of 1.20~ $M_{\odot}$ while Qian et al. (2006) found
1.1 $M_{\odot}$ from the use of a mass-period relation. These three values are consistent within the uncertainties.

The light curves of EV Cnc and ES Cnc show
asymmetries in their maxima. These can be due
to stellar activity on the surface of the stars.
We have proven that EV Cnc is a nearly contact binary system rather than a detached or an over-contact binary, consistent with
Blake's (2002) prediction. For the first time we presented the analysis of the complex light curve of EV Cnc. Different possibilities were considered during the LC solutions.
First of all, we computed solutions with different $q$-values  [0.30:0.95] and obtained the smallest sigma value at $q=0.5$.
The value of $q$ was chosen as adjustable parameter to perform the final solution.
The solution indicates a hot spot located on the surface of the star an impact region of flowing material from the secondary star   that faces the cooler component. Blake \& Rucinski (2004) reported a similar conclusion from their spectroscopic data.
We estimated the period variation rate from the O-C diagram to be 2$\times10^{-6}$ days per year.
This rate is higher than the one expected in semi-detached and contact binaries.
A reason for this could be the restricted available number of times of minima (spread over only $\sim$7 years) of the system.
New observations of the system are needed to estimate accurate mass accretion rate.
We could not decide firmly whether EV Cnc is a contact or a semi-detached binary.
The system can also be in a stage of evolution between these two configurations.
To obtain more accurate parameters,
the mass ratio should be estimated from high quality spectroscopic
observations.

We combined all the radial velocities of ES Cnc obtained by
van den Berg et al. (2001) and Sandquist et al. (2003) with the LCs obtained in this study.
We found $q= 0.777$ and $i=68^\circ$.3.
The solutions indicate that 18\% of the secondary star's surface is covered by cold spots. In addition,
the contribution of the third light, -previously spectroscopically
determined- at 0.25 phase was estimated to be 42\% ($l_3/{l_{total}}$) from
our observations. These results are in good agreement with those of
van den Berg et al. (2001) but differ from the values given by  Sandquist et al. (2003) and Pribulla et al. (2008).
We calculated the masses, radii, luminosities, etc. of the components of ES Cnc (Table~\ref{pyspar}).

Detached, semi-detached, and contact binaries belonging to a cluster,
provide us a precious laboratory to study the evolution of
close binary systems. There is an evolutionary connection between
late-type contact systems (LTCBs), semi detached (NCBs), and detached
(DCBs) systems (Yakut \& Eggleton 2005 and references therein for details).
The parameters of the close binaries AH Cnc, EV  Cnc, and ES Cnc are compared in Figures 1-3 of Yakut \& Eggleton (2005) with low-temperature contact, near-contact, and detached binaries whose physical parameters are well-known.
The figures show that they are in a good agreement. The evolution of an individual star in a close binary system is influenced by the
nuclear evolution, mass transfer (MT), mass loss (ML) and angular momentum loss (AML).
The effect of the AML mechanism depends on the mass of the component,
the mass ratio, and the separation of components. Using the obtained physical parameters
one can estimate the MT, ML, and AML  rates (see Yakut et al. 2008 for details)
for AH Cnc, EV Cnc, and ES Cnc.

Light and double-lined radial velocity curves of ES Cnc are available.  Using
the system parameters given in Table~\ref{pyspar}, we determined the
distance of the system and thus the one of the cluster to be 857(33) pc, while
the distance modulus is (m-M)$_V$=9.80(8).  The distance modulus given
previously by Sarajedini et al.\ (2004) amounts to 9.74(6), while Sandage et
al.\ (2003), Sandquist (2004), and Grocholski \& Sarajedini (2003) found 9.65,
9.72(5), and 9.64, respectively. The distance modulus we derived in this work is
thus in agreement with those previous studies.

\begin{acknowledgements}
    We thank the referee for providing constructive comments. KY acknowledges support by the Research Council of the University of Leuven under a DB fellowship and thanks the Turkish Scientific and Technical Research Council (T\"UB\.ITAK). WZ is supported by the FP6 European Coordination Action HELAS and by Research Council of the University of Leuven under grant GOA/2003/04.
    BK acknowledges support by the Communaut\'{e} Fran\c{c}aise de Belgique and Ministry of National Education, Turkey. SS is an Apirant Fellow of the Fund for Scientific Research, Flanders (FWO).
\end{acknowledgements}

\end{document}